# Low frequency non-resonant rectification in spin-diodes

R. Tomasello[1,*] B. Fang,[2,#] P. Artemchuk,[3,4] M. Carpentieri,[5] L. Fasano,[6] A. Giordano,[7] O. V. Prokopenko,[3] Z. M. Zeng,[2,*] G. Finocchio[7,*]

[1]Institute of Applied and Computational Mathematics, FORTH, GR-70013, Heraklion-Crete, Greece

[2]Key Laboratory of Multifunctional Nanomaterials and Smart Systems, Suzhou Institute of Nano-Tech and Nano-Bionics, CAS, Suzhou, Jiangsu 215123, People's Republic of China

[3]Faculty of RadioPhysics, Electronics and Computer Systems, Taras Shevchenko National University of Kyiv, Kyiv 01601, Ukraine

[4]Department of Physics, Oakland University, Rochester, MI 48309, USA

[5]Department of Electrical and Information Engineering, Politecnico of Bari, via Orabona 4, 70125 Bari, Italy

[6]Italian Space Agency (ASI), Italy,

[7]Department of Mathematical and Computer Sciences, Physical Sciences and Earth Sciences, University of Messina, I-98166, Messina, Italy

[*]corresponding authors: rtomasello@iacm.forth.gr, zmzeng2012@sinano.ac.cn, gfinocchio@unime.it

[#]now at Physical Science and Engineering Division, King Abdullah University of Science and Technology, Thuwal 23955-6900, Saudi Arabia

**Abstract**

Spin-diodes are usually resonant in nature (GHz frequency) and tuneable by magnetic field and bias current with performances, in terms of sensitivity and minimum detectable power, overcoming the semiconductor counterpart, i.e. Schottky diodes. Recently, spin diodes characterized by a low frequency detection (MHz frequency) have been proposed. Here, we show a strategy to design low frequency detectors based on magnetic tunnel junctions having the interfacial perpendicular anisotropy of the same order of the demagnetizing field out-of-plane component. Micromagnetic calculations show that to reach this detection regime a threshold input power has to be overcome and the phase shift between the oscillation magnetoresistive signal and the input radiofrequency current plays the key role in determining the value of the rectification voltage.

# I. INTRODUCTION

The recent demonstration that spin-transfer torque MRAMs (STT-MRAMs) based on magnetic tunnel junctions (MTJs) can be integrated with CMOS technology at chip level[1] has opened a concrete perspective for the development of hybrid spintronic/CMOS systems where the functionalities of the spin degree of freedom support the operations of standard electronics[2–4]. In addition to STT-MRAMs, MTJs spin-diodes have shown promising features for the integration in hybrid systems since better performances than CMOS counterpart, i.e. Schottky diode, have been achieved[5,6]. In particular, it has been experimentally proved that biased spin-diodes - working in active-regime – reach a sensitivity larger than 10kV/W, with output resistances smaller than 1kOhm[5,7–10]. This is possible thanks to the excitation of strongly nonlinear dynamics, such as nonadiabatic stochastic resonance[7], out-of-plane precession[11], nonlinear resonance[8], resonant vortex expulsion[9], as well as injection locking[5]. However, for energy saving in Internet-of-Things nodes and electromagnetic energy harvesting (background electromagnetic noise is characterized by a power density smaller than 10nW/cm$^2$), it is more convenient to design passive detectors that can work at ultralow power with a reasonably large signal-to-noise ratio[11,12]. Recently, it has been demonstrated that resonant spin-diodes can work at input power of nW level where Schottky diodes fail (see Fig. 6(a) in Ref. [[6]]). In order to use those diodes as electromagnetic energy harvesters, it is necessary to implement them as an array, where each spin-diode has a resonant frequency centered on a different value (e.g. DTV, GSM). However, spin-diodes can be also characterized by a non-resonant response[7,11]. Under certain conditions, they can work as broadband detectors, where a flat detection response in a wide range of input frequencies[6] is achieved. On the other hand, rectifications properties having a low frequency tail ($1/f$) for applications as MHz detectors have been recently observed[10,13]. In Ref. [[10]], the MTJ is designed to be superparamagnetic and, thanks to the adiabatic stochastic resonance, a high rectification voltage can be measured. In Ref. [[13]], the spin-diode is based on MTJs with interfacial perpendicular anisotropy (IPA) having a smooth linear resistance dependence with both the in-plane magnetic field and DC bias. This latter design scheme is promising for a low detection threshold.

In this work, we have experimentally studied the low frequency tail properties in unbiased MTJs with an engineered IPA near the compensation point where the IPA is of the same order of the out-of-plane component of the demagnetizing field and the second order anisotropy plays a role in the dynamical properties of the MTJ. Analytical calculations and micromagnetic simulations explain that the $1/f$ dependence of the rectified voltage is due to the reduction, as the frequency decreases, of the phase shift between the applied radiofrequency (RF) current and the oscillating magnetoresistance, while the amplitude of the magnetization oscillation is independent of the input frequency. In addition, some devices exhibit simultaneously low frequency broadband and high frequency resonant responses with a lower maximum voltage for a fixed input power. Our results show a rich variety of detector properties that can be engineered for more complex applications.

## II. EXPERIMENTAL DEVICE AND MEASUREMENTS

The spin-diodes are patterned into circular and elliptical nanopillars by using electron-beam lithography and ion milling techniques and are based on a different MgO-based MTJ stacks composed of PtMn (15)/$Co_{70}Fe_{30}$ (2.3)/Ru (0.85)/$Co_{40}Fe_{40}B_{20}$ (2.4)/MgO (0.8)/$Co_{20}Fe_{60}B_{20}$ (1.65) (thickness in nm), that, in term of material compositions, have been already optimized [14,15]. The thin films were deposited using a physical vapor deposition system and annealed at 300°C for two hours in a magnetic field of 1 T. The devices studied here are designed to have an in-plane pinned layer (PL) and a free layer (FL) with an effective anisotropy close to zero (near the point where demagnetizing field and IPA are totally compensated). The representative results shown in this work are for the elliptical device S1 with dimensions of 130 nm × 45 nm, while similar results have been obtained on a variety of samples with other dimensions S2-S5 (see details in the Table 1). We studied the spin-torque diode response by means of ferromagnetic resonance (FMR) measurements and by using the same setup as in Ref. [[5]]. All data were gathered at room temperature.

| Devices # | $H_{k1}$ (Oe) | $H_{k2}$ (Oe) | Size |
|---|---|---|---|
| S1 | 546 | 395 | 130 nm × 45 nm |
| S2 | 791 | 647 | 150 nm × 50 nm |
| S3 | 1088 | 912 | 100 nm × 100 nm |
| S4 | 1336 | 864 | 160 nm × 50 nm |
| S5 | 1395 | 1142 | 170 nm × 60 nm |

TABLE I: summary of the first and second order anisotropy constants and geometric characteristics of the devices investigated experimentally.

The junction resistance was measured as a function of an external in-plane magnetic field $H_{ext}$, as shown in the main panel of Fig. 1(a). $H_{ext}$ is applied parallel to the magnetic easy axis of the MTJ. When $H_{ext}$ increases from negative to positive, the FL magnetization gradually aligns from parallel towards antiparallel to the PL magnetization. The in-plane tunnel magnetoresistance ratio (TMR), defined as ($R_{AP}$-$R_P$)/$R_P$, is 84% where the resistances in the anti-parallel ($R_{AP}$) and parallel ($R_P$) configurations are 1660 Ω and 900 Ω, respectively. The resistance-area product in the parallel magnetization configuration was 4.2 Ω•μm². We estimate the angle between the magnetization vectors of the FL and the PL from the MTJ resistance to be 35° at zero bias field. The shift in the in-plane magnetoresistance-field scan is induced by the dipolar coupling from the polarizer.

Next, we measured the microwave rectified response of the device under a RF current. The RF current is applied to the device through a bias Tee using a signal generator (N5183B, Keysight). The precessional motion of the FL magnetization results in a time-dependent resistance oscillation, further producing a DC voltage across the MTJs, which is recorded with a Nanovolt meter (K2182, Keithley). As shown in Fig. 1(b), the microwave response spectra were extracted from the device under a RF power of 50 µW at low frequency regime. The FMR data clearly show that the amplitude of rectified voltage $V_{dc}$ can be modulated significantly through the external magnetic field, and, in particular, a low frequency detection occurs at the field range where the external field is of the same order of the dipolar field arising from the PL (near a zero effective bias field). Specifically, we focus on the data achieved for $H_{ext}$=90 Oe and we measure the rectified voltage as a function of frequency from 5 MHz to 2500 MHz with different RF input powers, as shown in Figs. 1(c) and (d). We can see that the magnitude of the $V_{dc}$ at the low frequency (5 MHz - 20 MHz) increases with increasing the input microwave power, consistently with the spectrum in Fig. 1(b), and with decreasing the input frequency (low frequency tail). The typical values of sensitivity (rectified voltage over input microwave power) and conversion efficiency (output power over input power) are larger than 100mV/mW and 0.1%, respectively. For example, at 5 MHz, they are 143 mV/mW and 0.2%, respectively. In order to investigate the physical origin of the low frequency tail, we have performed full micromagnetic simulations and developed an analytical theory to capture the qualitative features of the dynamics.

### III. MICROMAGNETIC MODEL AND PARAMETERS

The micromagnetic model is based on the numerical solution of the Landau-Lifshitz-Gilbert-Slonczewski equation by applying the time solver scheme Adams-Bashforth:

$$\frac{d\mathbf{m}}{d\tau} = -\left(\mathbf{m} \times \mathbf{b}_{eff}\right) + \alpha \left(\mathbf{m} \times \frac{d\mathbf{m}}{d\tau}\right) - \sigma \left[\mathbf{m} \times \left(\mathbf{m} \times \mathbf{m}_{\mathbf{p}}\right)\right] \quad (1)$$

where $\mathbf{m} = \mathbf{M}/M_s$ is the normalized magnetization of the MTJ FL, and $\tau = \gamma M_s t$ is the dimensionless time, with $\gamma$ being the gyromagnetic ratio, and $M_s$ the saturation magnetization. $\mathbf{b}_{eff} = \mathbf{B}_{eff}/M_s$ is the normalized effective field, which includes the exchange, magnetostatic, anisotropy, external, and thermal fields. $\alpha$ is the Gilbert damping. $\sigma = \sigma_\perp / \left(1 + \eta^2 \cos\beta\right)$ [16,17] is the current-torque proportionality coefficient, where $\eta = 0.66$ is the spin polarization factor and $\beta = \arccos(\mathbf{m} \cdot \mathbf{m}_{\mathbf{p}})$ is the angle between the FL's magnetization and PL's magnetization $\mathbf{m}_{\mathbf{p}}$ which is assumed to be completely fixed along the *x*-direction ($\mathbf{m_p} = \hat{\mathbf{x}}$, with $\hat{\mathbf{x}}$ being the unit vector of *x* axis). $\sigma_\perp = \frac{2\eta g \mu_B j_P}{2\gamma e M_S^2 t_{FL}}$ where *g* is the Landé factor, $\mu_B$ is the Bohr magneton, *e* is the electron charge, $t_{FL}$ is the FL thickness. $j_P$ is the RF current density flowing in the *z*-direction, $j_P = \frac{I(t)}{A_{FL}} = \frac{I_{ac} \cos(\omega t + \varphi)}{A_{FL}} = J_{ac} \cos(2\pi f_{AC} t + \varphi)$, where $I_{ac}$ is the magnitude of the input RF current, $I_{ac} \simeq \sqrt{2 P_{ac}/R_0}$, with $P_{ac}$ being the input RF power, and $R_0$ being the dc MTJ resistance. $A_{FL}$ is the FL area, $J_{ac}$ and $f_{ac}$ are the amplitude and frequency of the current, respectively, and $\varphi$ is the phase shift between the RF current and RF resistance oscillations.

The FL has an IPA of the first and second order, characterized by the anisotropy coefficients $k_1$ and $k_2$, that are computed from the experimental anisotropy fields shown in Table 1.

The thermal fluctuations at room temperature *T*=300 K are accounted by means of a stochastic field added to the deterministic effective field in each computational cell $\mathbf{b}_{th} = (\chi/M_S)\sqrt{2(\alpha K_B T / \mu_0 \gamma \Delta V M_s \Delta t)}$, with $K_B$ being the Boltzmann constant, $\mu_0$ is the vacuum permeability, *ΔV* the volume of the computational cubic cell, *Δt* the simulation time step, and $\chi$ a three-dimensional white Gaussian noise with zero mean and unit variance[18,19]. The noise is assumed to be uncorrelated for each computational cell. In the micromagnetic simulations and analytical calculations of this work, we consider the following parameters for the CoFeB FL[6]: $M_S$=950 kA/m, exchange constant *A*=20 pJ/m, $k_1$=0.48 MJ/m$^3$, $k_2$=1.5 ×10$^4$ J/m$^3$, $\alpha$ =0.02. In addition, only for the micromagnetic simulations, we investigate an elliptical 130nm × 46nm MTJ with $t_{FL}$=1.65 nm, as the experimental device S1, because it exhibits better performances. We use a discretization cell of 2.0x2.0x1.65 nm$^3$.

## IV. ANALYTICAL MODEL FOR THE RECTIFIED VOLTAGE

We develop an analytical model to capture the features of the low frequency tail dynamics and describe the rectified voltage as a function of the physical parameters. The main hypotheses are: (*i*) the magnetization of the FL **M** is spatially uniform and depends on time *t* only (i.e. macrospin approximation), *(ii)* a circular section for the spin-diode FL, as the device S3 in Table I, and *(iii)* the weak DC magnetic field $H_{ext}$ added to compensate the dipolar field deriving from the PL magnetization has a negligible impact on the magnetization dynamics, thus $\mathbf{B}_{eff}$ has contributions only from the demagnetization field $-\hat{\mathbf{z}}\mu_0 M_s m_z$ and IPA field $\hat{\mathbf{z}}\mu_0 \left[ H_{k1} m_z + s_2 H_{k2} m_z (1 - m_z^2) \right]$ [5]. In these expressions, $m_z = (\mathbf{m} \cdot \hat{\mathbf{z}})$ is the z-component of the vector $\mathbf{m}$, $H_{k1}$ and $H_{k2}$ are the first- and second-order anisotropy fields, respectively, $s_2 = \pm 1$ is the sign factor discussed below, and $\hat{\mathbf{z}}$ is the unit vector of the z axis. The field $\mathbf{B}_{\text{eff}}$ can be re-written in the form:

$$\mathbf{B}_{\text{eff}} = \hat{\mathbf{z}} \mu_0 \left[ s_0 h_0 + s_2 H_{k2} \left( 1 - m_z^2 \right) \right] m_z, \tag{2}$$

where we introduce the small field $h_0 = \mu_0 (H_{k1} - M_s)$, where the sign is defined by the factor $s_0 = \pm 1$.

Depending on the sign factors $s_0$ and $s_2$ in Eq. (2), several possible equilibrium states of the magnetization can exist in the considered spin-diode. Using the well-known relation $\mathbf{B}_{\text{eff}} = -(1/V) \partial E / \partial \mathbf{M}$ [20] ($V = A_{FL} \cdot t_{FL}$ is the FL volume), one can obtain the dependence of the magnetization energy $E$ on $m_z$ and find that the equilibrium cone state of the magnetization ($0 < |m_z| < 1$), observed in the experiment, is possible only when $s_0 = -1$ and $s_2 = +1$. Thus,

$$\mathbf{B}_{\text{eff}} = \hat{\mathbf{z}} \mu_0 \left[ -h_0 + H_{k2} \left( 1 - m_z^2 \right) \right] m_z.$$

We present the unit vector $\mathbf{m}$ in the form $\mathbf{m} = \hat{\mathbf{x}} \sin\theta \cos\phi + \hat{\mathbf{y}} \sin\theta \sin\phi + \hat{\mathbf{z}} \cos\theta$, where $\theta = \theta(t)$ is the out-of-plane (precession) angle between $\mathbf{m}$ and z-axis, and $\phi \equiv \phi(t)$ is the in-plane angle between the magnetization projection on $\hat{\mathbf{x}} - \hat{\mathbf{y}}$ plane and $\hat{\mathbf{x}}$ axis, $\hat{\mathbf{x}}$, $\hat{\mathbf{y}}$, $\hat{\mathbf{z}}$ are the unit vectors of x, y, z axes. By substituting $\mathbf{m}$ in Eq. (1), one can obtain the equations for $d\theta/dt$ and $d\phi/dt$:

$$\frac{d\theta}{dt} = -\frac{\alpha \omega_p \sin\theta + \sigma I(t)(\alpha \sin\phi + \cos\theta \cos\phi)}{1 + \alpha^2}, \tag{3a}$$

$$\frac{d\phi}{dt} = \frac{\omega_p - \sigma I(t)(\alpha \cot\theta \cos\phi - \csc\theta \sin\phi)}{1 + \alpha^2}, \tag{3b}$$

where $\omega_p \equiv \omega_p(\theta) = (\omega_2 \sin^2\theta - \omega_0)\cos\theta$ is the precession frequency, $\omega_0 = \gamma \mu_0 h_0$, $\omega_2 = \gamma \mu_0 H_{k2}$.

In the following, we make several additional assumptions. First, we assume that the damping is rather small, so one can neglect terms proportional to $\alpha^2$. Second, we assume that a weak input RF current excites magnetization oscillations near the equilibrium cone state of magnetization. In this regime of magnetization dynamics, the magnetization vector moves along an out-of-plane trajectory with the precession angle $\theta$ that is close to the equilibrium angle of the cone state of magnetization. Third, we assume that, at large input RF powers, $P_{ac} > P_{th}$, the input signal causes the transition of the magnetization dynamics to the mode where the magnetization vector moves along a circular out-of-plane trajectory with a very large out-of-plane angle $\theta \approx \pi/2$. We also consider the precession of the FL's magnetization along a circular orbit with an almost constant out-of-plane angle $\theta$, and time-dependent in-plane angle $\phi \approx s\omega t + \psi$, where $\psi$ is the phase shift between the magnetization precession and the RF input current. The factor $s = \pm 1$ defines the direction of the magnetization rotation. Under the previous hypotheses, the average of Eqs. (3) over the period of the RF current oscillations $2\pi/\omega$, by assuming $\langle d\theta/dt \rangle = 0$, $\langle d\psi/dt \rangle = 0$ ($\langle ... \rangle$ denotes averaging over the period of oscillations of the RF current), allows one to get the characteristic equation for the precession angle $\theta$:

$$(\omega_p - s\omega)^2 \sin^2\theta + \frac{u^2}{v^2}\alpha^2\omega_p^2 \tan^2\theta = \frac{u^2}{4}\sigma_\perp^2 I_{ac}^2, \tag{4}$$

where $u = \left[1 - \left(\sqrt{1-\eta^2}-1\right)^2/\eta^2\right]/\sqrt{1-\eta^2}$, $v = \left[1 + \left(\sqrt{1-\eta^2}-1\right)^2/\eta^2\right]/\sqrt{1-\eta^2}$.

It follows from the averaged equation for the slow variable $\psi$, $\langle d\psi/dt \rangle = -s\omega + 0.5 u \sigma_\perp I_{ac} \sin\psi$, that, at $\theta \approx \pi/2$, the stable magnetization dynamics is possible only if an input RF signal is rather large ($I_{ac} > I_{th}$ or $P_{ac} > P_{th}$), and the RF signal frequency is rather low ($f_{ac} < f_{th}$), where

$$I_{th} = \frac{2}{u}\frac{\omega}{\sigma_\perp}, \quad f_{th} = \frac{u}{4\pi}\sigma_\perp I_{ac}. \tag{5}$$

These equations give good approximations for the threshold current $I_{th}$ and frequency $f_{th}$ that are in agreement with the experimental data only when the input signal power is large, but the signal frequency is small. Otherwise, a more precise value of $I_{th}$ should be found numerically from Eq. (4): for a given input signal frequency $s\omega$, one can calculate $I_{ac}(\theta)$, and the minimum possible value of $I_{ac}(\theta)$ would be the frequency-dependent threshold $I_{th}(s\omega)$.

The output DC voltage generated by the spin-diode can be evaluated as $V_{dc} = \langle I(t) R(t) \rangle$, where $R(t) = R_\perp/(1 + \eta^2 \cos\beta)$ is the MTJ resistance, $R_\perp$ is the junction resistance in the perpendicular magnetic state, when the angle between $\mathbf{m}$ and $\mathbf{p}$ is $\beta_\perp = \pi/2$, $\beta \equiv \beta(t) = \arccos(\mathbf{m} \cdot \mathbf{p})$. Using expression for $\mathbf{m}$, the rectified voltage can be written in a form

$$V_{dc} = I_{ac} R_\perp \left\langle \frac{\cos(\omega t)}{1 + \eta^2 \sin\theta \cos\phi} \right\rangle \approx -w I_{ac} R_\perp \sqrt{1 - \left(\frac{I_{th}}{I_{ac}}\right)^2}, \tag{6}$$

where $w(\eta) = \left(1 - \sqrt{1-\eta^2}\right)/\eta\sqrt{1-\eta^2}$.

It follows from Eq. (6) that, above the threshold $P_{ac} > P_{th}$, the $V_{dc}$ should increase with an increase of the input RF power $P_{ac}$: $V_{dc} \sim \sqrt{I_{ac}^2 - I_{th}^2} \sim \sqrt{P_{ac} - P_{th}}$, where $P_{th} = 0.5 I_{th}^2 R_0$. However, with an increase of the input signal frequency $\omega$, the threshold current $I_{th}$ increases as well and, in turn, can substantially reduce the detector's output DC voltage, when $P_{ac}$ becomes comparable to the threshold power $P_{th}$ needed for the excitation of the low-frequency tail regime. This behavior of power- and frequency-dependences of $V_{dc}$ can been seen in Fig. 2. We also should notice again that Eq. (5) gives rather rough approximation for $I_{th}$ (it is precise enough only at $P_{ac} \gg P_{th}$, $f_{ac} \ll f_{th}$). Because of this, in order to provide a more accurate description of $V_{dc}(P_{ac})$ and $V_{dc}(f)$ curves shown in Fig. 2, we estimate $V_{dc}$ by substituting in Eq. (6) the threshold current $I_{th}$ calculated numerically from the characteristic Eq. (4), where $I_{th}$ corresponds to the minimum of the left-hand side of the equation.

## V. MICROMAGNETIC SIMULATIONS RESULTS

Figure 3(a) shows the output DC voltage $V_{dc}$ in the low frequency regime for two values of the current density amplitude, which can be related to low and high input powers, respectively. The $V_{dc}$ is calculated by $V_{dc} = 0.5 I_{ac} \Delta R_s \cos(\psi)$, where $\Delta R_s$ is the amplitude of the oscillating resistance estimated directly from the simulations by considering the experimental values of $R_{AP} = 1660 \ \Omega$ and $R_P = 900 \ \Omega$. The magnitude of $V_{dc}$ shows a decreasing monotonic behavior as the frequency increases, in qualitative agreement with the experimental outcomes. In order to understand the origin of such a behavior, we need to refer to the analytical expression of $V_{dc}$, specifically to the time dependences of the oscillating resistance and phase shift. Figure 3(b) and (c) depict the time domain evolution of the x-component of the normalized magnetization together with the RF normalized current density $j_{ac}$, for $J_{ac} = 2.0 \ \text{MA/cm}^2$ when $f_{ac}$=10 MHz and 100 MHz, respectively. The amplitude of the oscillating magnetization is maximum regardless the frequency, while a difference phase shift can be noticed for the two frequencies (similar achievements are obtained for $J_{ac} = 0.8 \ \text{MA/cm}^2$). Therefore, we calculate the phase shift as a function of the frequency and we plot the results in Fig. 3(d). We observe that the phase shift increases in magnitude as the frequency increases, thus proving its key role in reducing the rectified voltage. The simulation results are robust over a range of parameters (not shown), hence we can conclude that the low frequency tail measured in the experiments is due to a change of the phase shift between the applied RF current and the resulting oscillating resistance.

## VI. SUMMARY AND CONLUSIONS

In summary, we have experimentally investigated spin-diodes characterized by an IPA near the compensation point where it balances the demagnetizing field out-of-plane component and by a non-negligible second order anisotropy. We have shown that, for a certain power range, the broadband response is converted into a low frequency tail with a $1/f$ dependence of the rectified voltage. We have developed an analytical theory and performed full micromagnetic simulations to qualitatively comprehend the spin-diode magnetization dynamics and understand the origin of the low frequency tail. We have demonstrated that the amplitude of the oscillation is independent of the frequency, while the phase shift between the applied RF current and the oscillating magnetoresistance is the key factor in giving rise to the low frequency tail. In addition, since this low frequency tail is observed when the dipolar field from the PL is compensated by the external field, we believe that the use of PL with reduced dipolar field (less than 2mT) can be the basis for the design of zero field low frequency detectors. Finally, we wish to highlight that the variety of frequency response exhibited by the spin-diodes can be very useful to design novel applications beyond the detection.

## ACKNOWLEDGMENTS

This work was supported under the Grant 2019-1-U.0. ("Diodi spintronici rad-hard ad elevata sensitività - DIOSPIN") funded by the Italian Space Agency (ASI) within the call "Nuove idee per la componentistica spaziale del futuro" and the Executive Programme of Scientific and Technological Cooperation Between Italy and China (2016YFE0104100). R.T. and G.F. also acknowledge the project ThunderSKY, funded by the Hellenic Foundation for Research and Innovation (HFRI) and the General Secretariat for Research and Technology (GSRT), under Grant Agreement No. 871. P.A. and O.V.P. acknowledge the Grants Nos. 18BF052-01M and 19BF052-01 from the Ministry of Education and Science of Ukraine and Grant No. 1F from the National Academy of Sciences of Ukraine. This work was also supported by PETASPIN association.

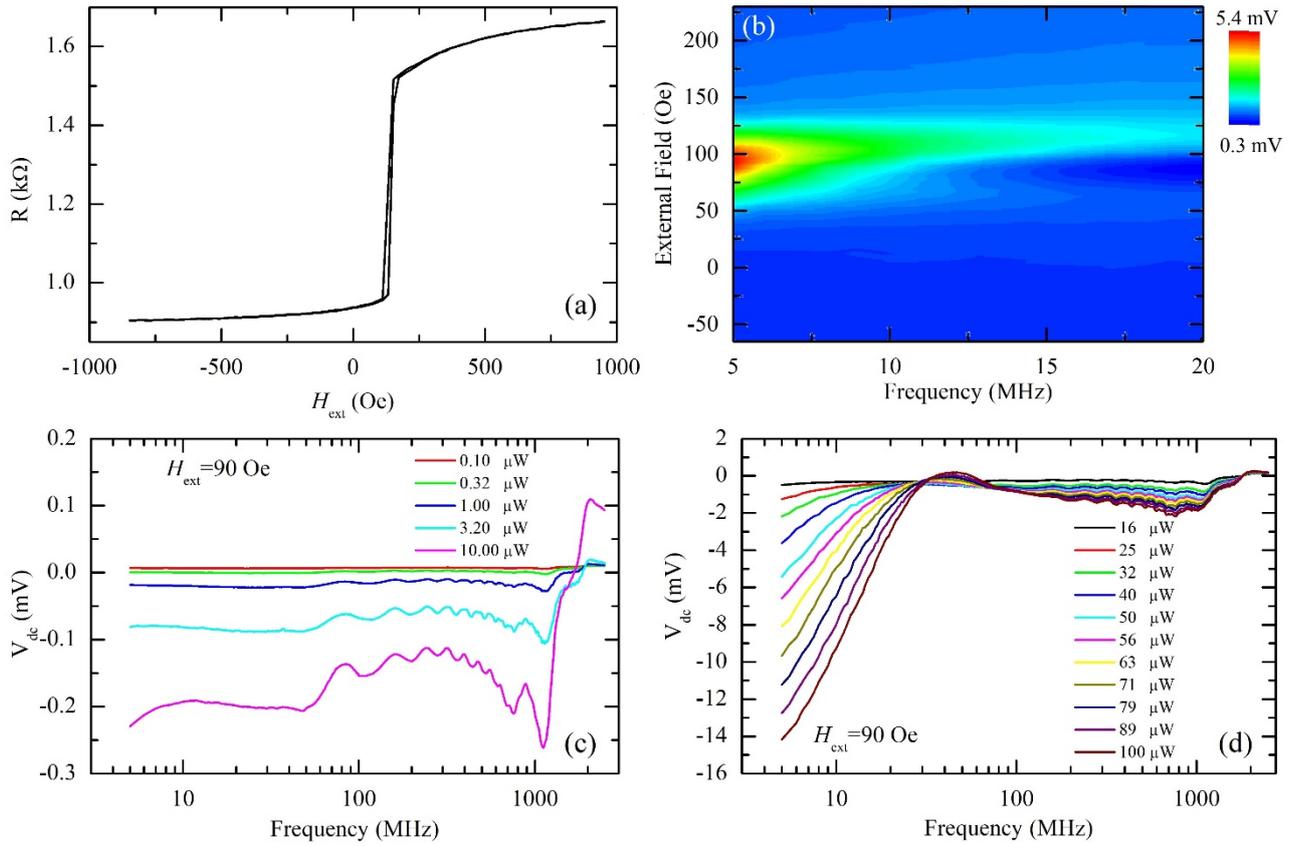

Fig. 1. (a) The magnetoresistance curve of the MTJ device under in-plane magnetic fields $H_{ext}$ for $I_{dc}$ = 10 μA, showing a sectional perpendicular FL magnetization. (b) Microwave rectification response spectra under RF power of 50 μW at low frequency regime (from 5 MHz to 20 MHz). (c) and (d) Rectified voltage ($V_{dc}$) as a function of RF frequency for various RF powers ($P_{ac}$) at $H_{ext}$ = 90 Oe.

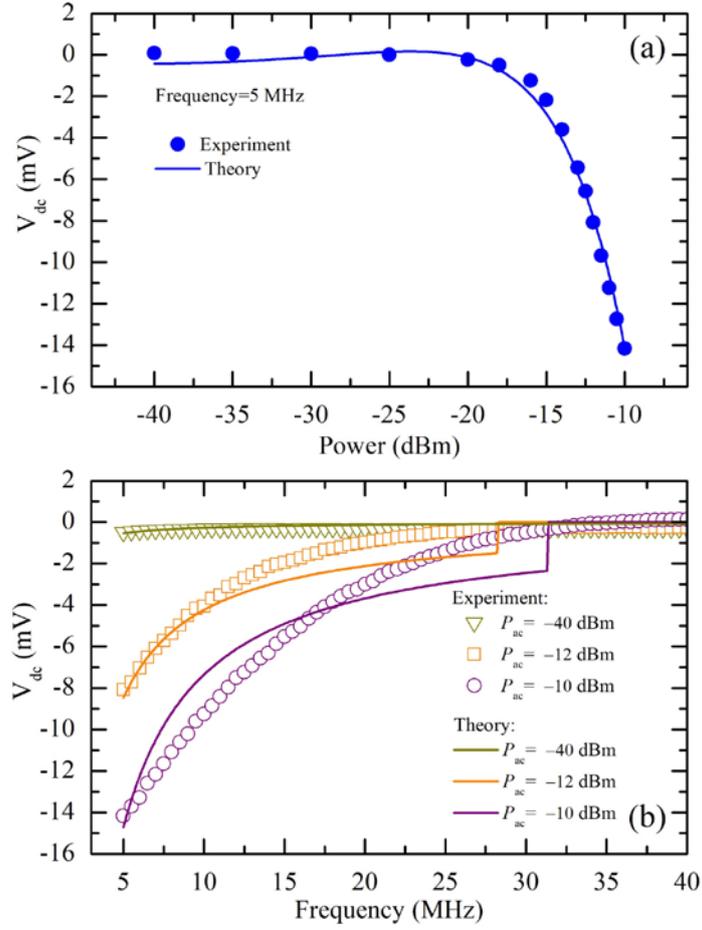

Fig. 2: (a) Dependence of the spintronic diode's output DC voltage $V_{dc}$ on the input signal power $P_{ac}$ for the signal frequency of 5 MHz. (b) Frequency dependences of the diode's output DC voltage $V_{dc}$ for the different input signal powers $P_{ac}$. In both panels, the symbols represent the experimental data, while the solid lines indicate the analytical results.

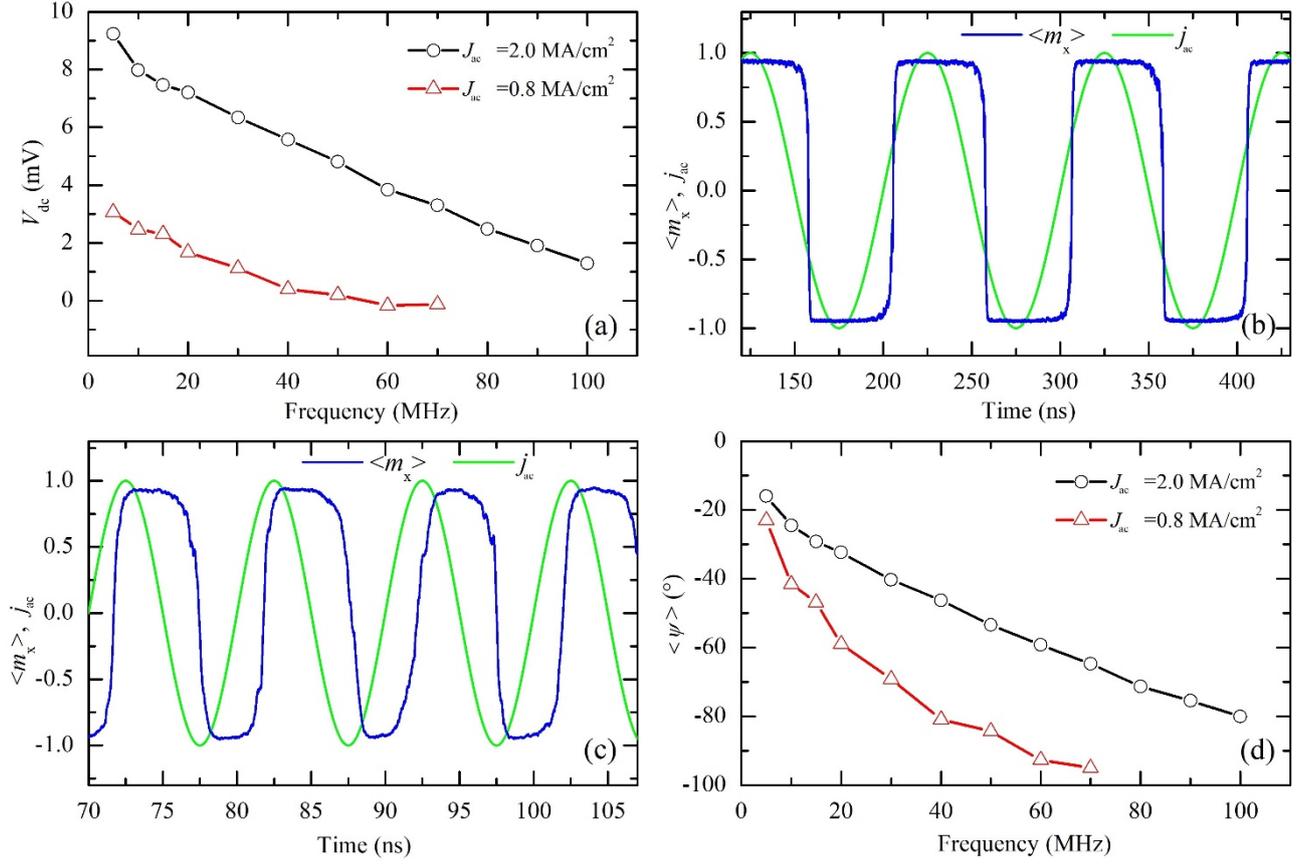

Fig. 3 Micromagnetic simulations results (a) Output DC voltage as a function of the applied current frequency for two values of the current density amplitude. Time evolution of the normalized *x*-component of the magnetization (blue curve) and normalized current density (green curve) for $J_{ac} = 2.0$ MA/cm$^2$ and (b) $f_{ac} = 10$ MHz, and (c) $f_{ac} = 100$ MHz. (d) Time-averaged phase shift as a function of the applied current frequency for two values of the current density amplitude.